\documentclass[smus]{snow2e}
\usepackage{graphics}

\begin{document}

\title{Single Neutral Heavy Lepton Production at $e^{+}$ $e^{-}$
and $\mu^{+}$ $\mu^{-}$ Colliders\thanks{Work supported in part by the
Natural Sciences and Engineering Research Council of Canada}}

\author{Pat Kalyniak and I. Melo\\ {\it Ottawa Carleton Institute for
Physics}\\ {\it Department of Physics, Carleton University}\\
{\it 1125 Colonel
By
Drive, Ottawa, Ontario, Canada K1S 5B6}}

\maketitle

\thispagestyle{empty}\pagestyle{empty}

\begin{abstract} 
Many extensions of the Standard Model include extra fermions. We here
consider a model containing Dirac neutral heavy leptons in addition to
massless neutrinos. These neutral heavy leptons can be produced singly
along with a massless neutrino in $e^{+}$ $e^{-}$ and $\mu^{+}$ $\mu^{-}$
collisions. The production rates depend on the neutral heavy lepton
masses and mixing parameters. We here consider the rates for this process
and potential signatures for the machine energies and luminosities
suggested for the Snowmass study. 
\end{abstract}

\section{Introduction and Model}
We have previously considered ~\cite{melo} a simple extension of
the standard model (SM) with an enriched neutral fermion spectrum
consisting a of massless neutrino and a Dirac neutral heavy lepton (NHL)
associated with each generation ~\cite{vallemo,bernabeu1,wolfe}. In
$e^{+}$
$e^{-}$ and $\mu^{+}$ $\mu^{-}$ colliders,these NHL's can be pair produced
or produced singly in association with a massless neutrino ~\cite{buch,azdj,finns}. The kinematic
reach in NHL mass of the latter process exceeds that of the former so we
consider only the single NHL production here.

The model which we consider here includes
two new weak isosinglet neutrino fields per generation.
Imposing total lepton number conservation and diagonalizing the mass matrix
yields three massless
 neutrinos ($\nu_{i}$) along with three Dirac NHL's ($N_{a}$) of mass
 $M_{N}$. The weak interaction eigenstates ($\nu_{l}, l=e,\mu,\tau$)
are related to the six mass eigenstates
via a $3 \times 6$ mixing matrix $K \equiv (K_{L}, K_{H})$:
\begin{eqnarray}
\label{eq:alter}
\nu_{l} & = & \sum_{i=1,2,3}\big(K_{L}\big)_{li} \nu_{i_{L}}
+ \sum_{a=4,5,6} \big(K_{H}\big)_{la} N_{a_{L}}.
\end{eqnarray}
Various combinations of the mixing matrix parameters arise.
The mixing factor which typically governs flavour-conserving
processes, $ll_{mix}$,
is given by
\begin{eqnarray}
ll_{mix} & = & \sum_{a=4,5,6} \big(K_{H}\big)_{la}
\big(K_{H}^{\dagger}\big)_{al}\;\; ,
\makebox[.3in] [c] { } l= e, \mu, \tau
\end{eqnarray}
and the flavour-violating mixing factor   $l{l^{'}}_{mix}$
is defined as
\begin{eqnarray}
l{l^{'}}_{mix} & = & \sum_{a=4,5,6} \big(K_{H}\big)_{la}
\big(K_{H}^{\dagger}\big)_{al^{'}}\;\; ,
\makebox[.3in] [c] { }l \neq l^{'}.
\end{eqnarray}
Further, the following inequality holds
\begin{eqnarray}
\label{eq:ineq}
|{l{l^{'}}_{mix}}|^2 & \leq & {ll}_{mix}\:\:{l^{'}l^{'}}_{mix},
\makebox[.3in] [c] { } l \neq l^{'}.
\end{eqnarray}
We choose from the above the parameters $ee_{mix}$, $\mu \mu_{mix}$ and $\tau
\tau_{mix}$ to characterize the model, while neglecting the flavour
violating parameters, $e \mu_{mix}$, $e \tau_{mix}$ and $\mu \tau_{mix}$.
The inequality above implies this could be reasonable and, also, we
note that the parameter $e \mu_{mix}$ is the most tightly constrained by
experiment. The existing constraints are~\cite{ggjv}
 $|e\mu_{mix}| \leq  0.00024$ 
and~\cite{Nardi}
\begin{eqnarray}
\label{eq:limits1}
      ee_{mix} & \leq & 0.0071 \nonumber \\
      \mu\mu_{mix} & \leq  & 0.0014  \\
      \tau\tau_{mix} & \leq & 0.033 \nonumber
\end{eqnarray}
The best constraints on the mixings $e \tau_{mix}$ and $\mu \tau_{mix}$
arise from the inequality above and are not particularly strong.

We consider below the production, in $e^{+}$ $e^{-}$ and $\mu^{+}$ $\mu^{-}$
colliders, of a single NHL along with a massless neutrino, parametrized by
the NHL mass, $M_{N}$, and the mixings $ee_{mix}$, $\mu \mu_{mix}$ and $\tau
\tau_{mix}$. We discuss the NHL decay and, hence, the potential signatures
for the process; in particular we focus on the signature whereby the NHL
can be fully reconstructed. We estimate the limits on $M_{N}$ and the
mixings which can be reached for various machine parameters. We here
consider only NHL's with mass greater than $M_{W}$ and $M_{Z}$, given the
absence of experimental evidence for the production of NHL's as decay
products of $W$ and $Z$ bosons \cite{lang,verzeg}.


%
%

\section{Single Neutral Heavy Lepton Production}
\begin{figure}[h]
\leavevmode
\begin{center}
\resizebox{!}{3cm}{%
\includegraphics{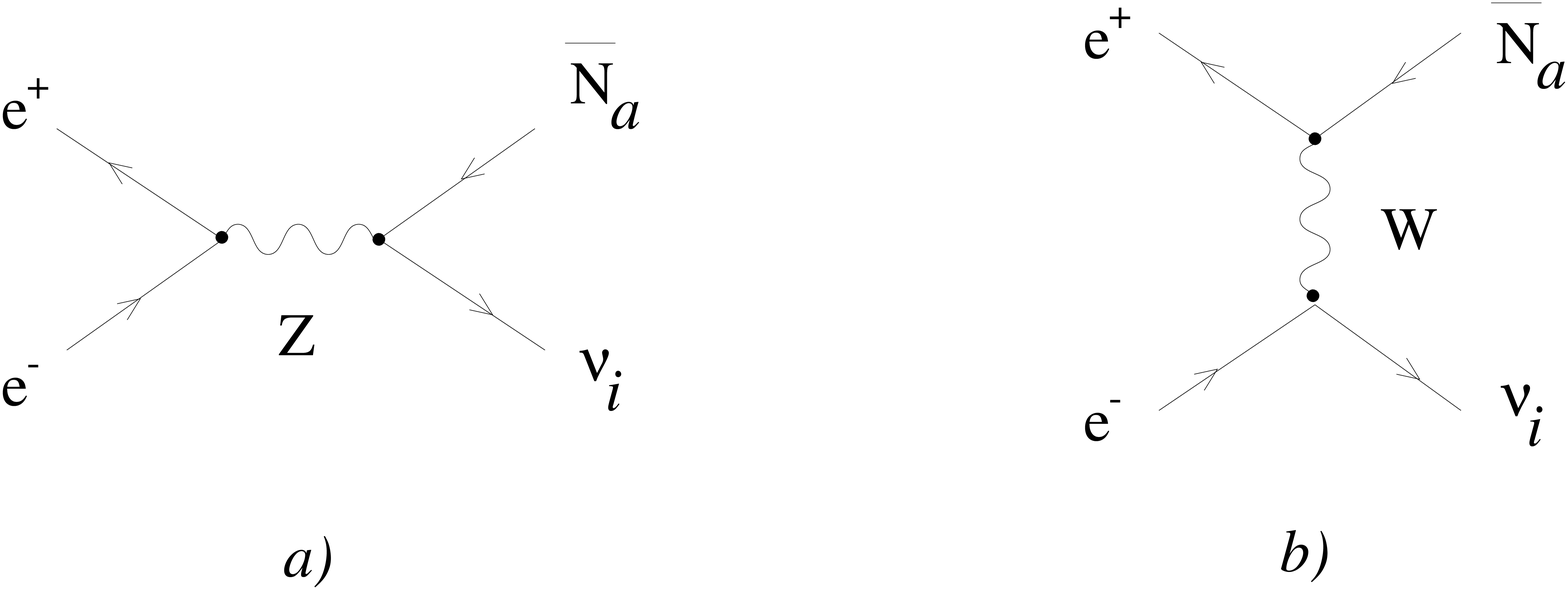}}
\end{center}
\caption{Feynman diagrams for $e^+e^- \rightarrow \overline{N_{a}} \nu_{i}$: a) s-channel, b) t-channel.}
\label{fig1}
\end{figure}

The production of a single NHL, $N_{a}$, in association with a massless
neutrino, $\nu_{i}$, proceeds via the two diagrams of Figure \ref{fig1}. 
In the
context of the model described in the last Section, this process is
determined by the following terms of the Lagrangian.

\begin{eqnarray}
\label{eq:ccur}
{\cal L} & = & \frac{1}{2 \sqrt{2}} g W^{\mu} \sum_{l=e, \mu ,\tau}
 \Big\{ \; \sum_{i} \bar
{l} \gamma_{\mu} (1-\gamma_{5}) \big(K_{L}\big)_{li} \nu_{i}
\nonumber \\
& + & \sum_{a} \bar{l} \gamma_{\mu} (1-\gamma_{5}) 
 \times \big(K_{H}\big)_{la} N_{a} \Big\}  \nonumber \\ 
 & + & \frac{g}{4c_{W}} Z^{\mu} \sum_{i,a}
 \bar{\nu_{i}}
{(K_{L}^{\dagger}K_{H})}_{ia} \gamma_{\mu} (1-\gamma_{5})N_{a}
+ h.c.
\end{eqnarray}

\noindent
The resulting cross section for single NHL production, 
$\sigma \equiv \sigma(l^{+}l^{-} \rightarrow N_{a}\bar{\nu}_{i})$, is
given below. 
\begin{eqnarray}
\label{eq:xsecn}
\sigma & = & 
\frac{\pi \alpha^{2}}{4 s_{W}^{4} s} \{t_{mix} [(1-\chi_{N})
(2+\chi_{W}^{-1}) - (2+2\chi_{W} \nonumber \\
& - & \chi_{N})\ln r] + s_{mix} \frac{a^2+b^2}{6 c_{W}^2} \frac{(1-\chi_{N})^2
(2+\chi_{N})}{(1-\chi_{Z})^2} \nonumber \\
& + & st_{mix} \frac{a}{c_{W}^2} \frac{1}{(1-\chi_{Z})}
[\frac{1}{2}(1-\chi_{N})(3+2\chi_{W}-\chi_{N}) \nonumber \\
& - & (1+\chi_{W})(1-\chi_{N}+\chi_{W})\ln r] \}
\end{eqnarray}
\noindent 
In the above, $\chi_{j} \equiv M_{j}^2/s$, $j = N, W, Z$ and 
$r \equiv \frac{1-\chi_{N}+\chi_{W}}{\chi_{W}}$. Also, $a = -
\frac{1}{2} + s_{W}^2$ and $b = s_{W}^2$.
The
coefficients of the t-channel $W$ exchange, the s- channel $Z$ exchange,
and the interference, namely, $t_{mix}$, $s_{mix}$, and $st_{mix}$, depend
on the entries in the mixing matrix $K \equiv (K_{L}, K_{H})$. In terms of
the parameters $ee_{mix}$, $\mu \mu_{mix}$ and $\tau
\tau_{mix}$, they are limited as follows.
\begin{eqnarray}
\label{eq:mix}
 t_{mix} & = & |(K_{L}^{*})_{li}(K_{H})_{la}|^{2}\; \leq \;
ll_{mix}(1-ll_{mix})
   \nonumber \\ 
 s_{mix}& = & |(K_{L}^{\dagger} K_{H})_{ia}|^{2} \; \leq \;
\sum_{l=e,\mu,\tau} ll_{mix} (1-ll_{mix})  \\
st_{mix} & = & (K_{L}^{\dagger} K_{H})_{ia} (K_{L})_{li}
(K_{H}^{*})_{la} \; \leq \; ll_{mix} (1-ll_{mix}) \nonumber 
\end{eqnarray}

\begin{figure}[b]
\leavevmode
\begin{center}
\resizebox{!}{11cm}{%
\includegraphics{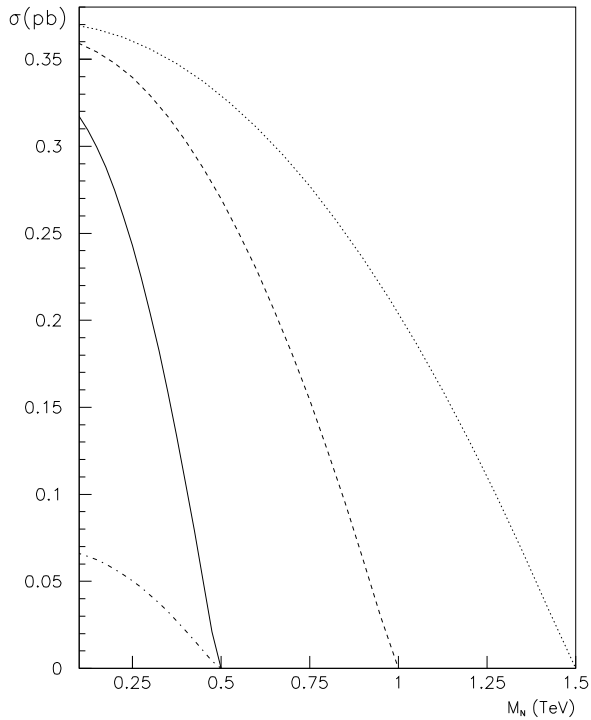}}
\end{center}
\caption{Total cross section vs NHL mass $M_{N}$ for $e^+e^-$ collider at three different energies: $\sqrt{s}=0.5$ TeV (solid line), $\sqrt{s}=1.0$ TeV (dashed line) and $\sqrt{s}=1.5$ TeV (dotted line) and for $\mu^+\mu^-$ collider at $\sqrt{s}=0.5$ TeV (dash-dotted line); $ee_{mix} = 0.0071$,  $\mu\mu_{mix} = 0.0014$, $\tau\tau_{mix} = 0.033$.}
\label{fig2}
\end{figure}

\begin{figure}[b]
\leavevmode
\begin{center}
\resizebox{!}{11cm}{%
\includegraphics{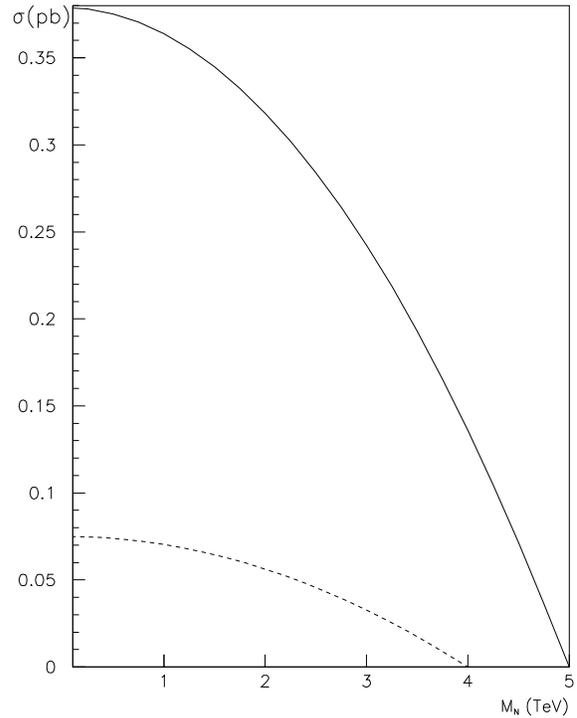}}
\end{center}
\caption{Total cross section vs NHL mass $M_{N}$ for $\sqrt{s}=5$ TeV $e^+e^-$ collider (solid line) and $\sqrt{s}=4.0$ TeV $\mu^+\mu^-$ collider (dashed line); $ee_{mix} = 0.0071$,  $\mu\mu_{mix} = 0.0014$, $\tau\tau_{mix} = 0.033$.}
\label{fig3}
\end{figure}

The single NHL production cross sections are displayed as a function of NHL
mass $M_{N}$ in Figures \ref{fig2},\ref{fig3} for the various machine parameters of the
Snowmass study. In Figure \ref{fig2}, the lower energy machine results are given.
These include $e^{+}$ $e^{-}$ colliders with $\sqrt{s} = 0.5$, 1.0, and 1.5
TeV and a $\mu^{+}$ $\mu^{-}$ collider of $\sqrt{s} = 0.5$ TeV. The
proposed integrated luminosities for these machines are 50, 200, and 200
$fb^{-1}$ for the three $e^+$ $e^-$ collider options, respectively, and
either 7 or 50 $fb^{-1}$ for the muon collider. In Figure \ref{fig3}, the
results for the high energy options, a 5 TeV $e^{+}$ $e^{-}$ collider and a
4 TeV $\mu^{+}$ $\mu^{-}$ collider, are displayed. For each of these
options the proposed integrated luminosity is 1000 $fb^{-1}$. Hence, one
can expect tens of thousands of signal events for a large range of $M_N$. 
In these Figures,
the mixing parameters used are those quoted in Equation~\ref{eq:limits1}
, namely, the present
experimental limits, so these results should be considered upper limits.
Note, in the $e^{+}$ $e^{-}$ case the coefficient
$ee_{mix} (1-ee_{mix})$ multiplies the t-channel and s-t channel
interference terms, whereas in the $\mu^{+}$ $\mu^{-}$ collider case, these
terms come in with the coefficient $\mu \mu_{mix} (1-\mu \mu_{mix})$. 
Note also that these results should be multiplied by a factor of two to
include the conjugate process as well, $\sigma(l^+ l^- \rightarrow 
N \bar{\nu} + \bar{N} \nu)$.

The t-channel $W$ exchange diagram of Figure \ref{fig1} is by far the dominant one
relative to the s-channel $Z$ diagram. Hence, the production cross section
for the $\mu^{+}$ $\mu^{-}$ collider scales approximately from that of the 
$e^{+}$ $e^{-}$ collider as 
$\frac{\mu \mu_{mix} (1-\mu \mu_{mix})}{ee_{mix} (1-ee_{mix})}$. Because of
this t-channel dominance any of our results can be easily scaled to smaller
mixing parameters, to a good approximation.Similarly, the overall $M_{N}$
dependent factor of this t-channel term, $[(1-\chi_{N})
(2+\chi_{W}^{-1}) - (2+2\chi_{W}-\chi_{N})\ln r]$
, provides the approximate
scaling with $M_{N}$.

\section{Signatures and Results}
We must consider the possible decay modes of the NHL \cite{pil}. For NHL mass large
relative to the $W$ and $Z$ masses, the NHL will decay into a light lepton
($\nu$, $l$) and a gauge boson ($W$, $Z$) which subsequently also decays.
If a Higgs boson is kinematically accessible, the NHL can also decay into a
massless neutrino along with the Higgs boson. Table \ref{tab:brs} shows the branching
ratios into these modes in the two cases that a $H^{0}$ is or is not
kinematically accessible. 
\begin{table*}[t]
\begin{center}
\caption{NHL decay branching ratios and production signatures}
\label{tab:brs}
\begin{tabular}{rccl}
\hline
\hline
Decay Mode  & B.R. with $H^{0}$ & B.R. no $H^{0}$ & Signatures \\
\hline
$N \rightarrow l W$  &   1/2 & 2/3 & $(l(jj)_{W})_{N}, \; ll'$  \\
$\rightarrow \nu Z$ &  1/4 & 1/3& $(jj)_{Z},\; (l^{+}l^{-})_{Z}$  \\
$\rightarrow\nu H$   &   1/4 & - & $(jj)_{H},\; ((jj)_{W}(jj)_{W})_{H}$ 
\\ \hline
\hline
\end{tabular}
\end{center}
\end{table*}
The resulting signatures for the single NHL
production are also shown in the Table; it should be understood that in
each case one also has missing energy and momentum due to the undetected
$\nu$('s) as part of the signature. The distribution among the particular
final
state lepton flavours is dictated by the details of the mixing parameters
in the matrix $K$, whereas it is the combinations $ll_{mix}$ and
$ll_{mix}'$ which are
constrained by experiment. Hence the signatures for the NHL production
should be understood to include all lepton flavours in an unspecified
distribution. The subscripts $W$, $Z$, and $H$
indicate the appropriate invariant mass of the observed particles. In the
case of the NHL decay $N \rightarrow l W$ with the subsequent hadronic
decay of the $W$, the NHL mass can be fully reconstructed. Hence we focus
on this signature. 

In an earlier study ~\cite{buch} on single NHL
production at a 500 $GeV$ $e^+$ $e^-$ collider, only events in which the
$W$
decayed leptonically were considered. In that case, the signature is a
pair of energetic charged leptons with a small opening angle. Backgrounds
include single $W$ and $W$ pair production, with the $W$'s decaying
leptonically. Therein,
a set of cuts which rejected the $W$ pair background at the level of about
99.4\% was implemented. It was also argued that the single $W$ production
was 
controllable. For an integrated luminosity of 20 $fb^{-1}$, they expected
sensitivity to NHL mass up to $M_{N} = 410 \;
GeV$ for a t-channel mixing factor of $t_{mix} = 0.01$ and for a 300 $GeV$
NHL, they estimated a sensitivity down to $t_{mix}$ of $1 \cdot 10^{-5}$.

Those results were encouraging. However, the use of the events in which
the $W$
decays hadronically to boost the statistics and also to reconstruct the
NHL mass will extend the possible discovery limits. The branching ratio
for NHL decay to this mode ranges from
$\frac{1}{2}$ to $\frac{2}{3}$ with an extra factor of $\frac{2}{3}$ for
the branching ratio of the $W$ into its hadronic decay mode. Backgrounds
competing with this lepton, two-jet, missing momentum signal include
single $W$ production followed by hadronic decay of the $W$, $W$ pair
production with one $W$ decaying hadronically and the other leptonically, 
and also the two photon
process $\gamma \gamma \rightarrow l^+ l^- q \bar{q}$, where one of the
leptons is undetected. Here also, there has been a study comparing this
signature and its backgrounds for the case of a 500 $GeV$ $e^+$ $e^-$
collider~\cite{azdj}, with $t_{mix} = s_{mix} = st_{mix} \equiv
(\theta_{mix})^2$. A set of cuts was determined which effectively
lowers the background rate while maintaining most of the signal; the
distribution in
$ljj$ invariant mass was then used to distinguish the $(ljj)_{N}$ peak
about the NHL mass
from the background. The cuts include demanding one and only one charged
lepton, which eliminates much of the single $W$ production background as
the accompanying charged lepton is likely to go along the beam pipe in
that process. Also, the $(jj)$ invariant mass must reconstruct to the $W$
mass, within some resolution.
Because the NHL can be fully reconstructed, so can the missing $\nu$
momentum. The remaining cuts, which are fully described in~\cite{azdj},
include constraints involving the missing momemtum and the reconstructed
NHL. In their example of $\theta_{mix} = 0.025$ and $M_{N}$  = 350
$GeV$,
while the signal rate is reduced by the cuts by a factor of 0.85, the
single $W$ rate is reduced by a factor of $1.7 \cdot 10^{-3}$ and the $W$
pair rate by a factor $3.5 \cdot 10^{-3}$. The two photon rate was reduced
to unobservability on imposition of the cuts. They find, for $\theta_{mix}
= 0.025$ ($t_{mix} = 6.25 \times 10^{-4}$), sensitivity up to $M_{N}$ = 450 $GeV$.This particular result is
based on about 175 events (before B.R.'s and cuts) for an integrated
luminosity of 20 $fb^{-1}$. For $M_{N}$ of 350 $GeV$,
they estimate sensitivity to a mixing down to $\theta_{mix}$ of about
0.005 ($t_{mix} = 2.5 \cdot 10^{-5}$). This result is based on just over
20 events for the same luminosity.

We do not present a detailed comparison of the NHL signal and its
backgrounds here. Rather we make some estimates based on the results
quoted above for the case of the 500 $GeV$ $e^+$ $e^-$ collider. The cross
section for single NHL production asymptotically approaches a constant
value with increasing $s$. On the other hand, beyond threshold, the $W$
pair production cross section falls as $s$ increases. The kinematic nature
of these processes does not change with increasing $s$ so the cuts
implemented in~\cite{azdj} will remain effective. Hence the $W$ pair
background will not be a problem for $\sqrt{s}$ beyond 500 $GeV$.
Similarly, the two photon process is under control. However, the single
$W$ production ($l^+ l^- \rightarrow l \nu W$) cross section also rises
asymptotically to a constant value as $s$ increases. Hence, this process
remains the limiting background for the higher energy colliders but
the cuts should be effective in reducing this source of background. We
assume, then, that the backgrounds can be controlled. This must be
confirmed by detailed study and so we present only estimates of discovery
limits.


Given that the detailed study of Azuelos and Djouadi~\cite{azdj} based
their discovery limits on  rates, before B.R.'s and cuts, of the order
of 100 events, we use this number of events to determine our discovery
limits for the higher energy machines. If we assume the mixings
 take the values in Eq.~\ref{eq:limits1}, the reach in NHL mass approaches
the kinematic limit for each of the high energy colliders. For some other
representative values of NHL
mass, we show in Table \ref{tab:disclim}, approximate limits
in mixing $t_{mix}$, for the case $t_{mix} = s_{mix} = st_{mix}$, for 
the various machine energies and proposed integrated luminosities. 
These limits are based on the cross section (multiplied by a factor of 2 to include the conjugate process) given in Eq. \ref{eq:xsecn}. Using the t-channel dominance and $s, M_{N} >> M_{W}$, we can further simplify Eq. \ref{eq:xsecn} to obtain the following approximation for $t_{mix}$:
\begin{eqnarray}
\label{eq:tmixa}
t_{mix}& = &\frac{1.44\;\: s(TeV^2)\;\: n}{L(pb^{-1})\;f(s,M_N)}, \\
f(s,M_N)& = & (1-\chi_N)\chi_W^{-1} - (2-\chi_N)\ln{\frac{s-M_N^2}{M_W^2}}
\nonumber
\end{eqnarray}
where $n$ is the number of events (100) and $L$ is the integrated luminosity. For a large interval of $M_{N}$, also the logarithmic term of $f(s,M_N)$ can be neglected. 

The results of Table \ref{tab:disclim} thus indicate that the
higher energy colliders will be sensitive to mixings in the range $1 \cdot
10^{-5}$ down to $1 \cdot 10^{-6}$ over a large range of NHL masses,
$M_{N}$.  

\begin{table}[t]
\begin{center}
\caption{Discovery limits for NHL masses and mixings}
\label{tab:disclim}
\begin{tabular}{llll}
\hline
\hline
$\sqrt{s}$ ($TeV$)  & $\int dt \cal{L} $ ($fb^{-1}$) & $M_{N}$ ($TeV$) &
$t_{mix}$ \\
\hline
1.0  &   200 & 0.5 & $7 \cdot 10^{-6}$  \\
 & & 0.75 & $1 \cdot 10^{-5}$ \\
 & & 0.95 & $6 \cdot 10^{-5}$ \\
1.5 & 200 & 0.5 & $5 \cdot 10^{-6}$ \\
 &  & 1.0 & $9 \cdot 10^{-6}$ \\
 &  & 1.25 & $2 \cdot 10^{-5}$ \\
 &  & 1.45 & $8 \cdot 10^{-5}$ \\
4.0 & 1000 & 0.5 & $9.5 \cdot 10^{-7}$ \\
 &  & 1.0 & $1 \cdot 10^{-6}$ \\
 &  & 2.0 & $1.2 \cdot 10^{-6}$ \\
5.0 &  & 0.5 & $9.4 \cdot 10^{-7}$ \\
 &  & 1.0 & $9.7 \cdot 10^{-7}$ \\
 & & 2.0 & $1.1 \cdot 10^{-6}$ \\
\\ \hline
\hline
\end{tabular}
\end{center}
\end{table}

\section{Conclusions}

We have investigated the production in $e^+$ $e^-$ and $\mu^+$ $\mu^-$
colliders of a single NHL along with a massless neutrino, for collider
energies ranging from 500 $GeV$ to 5 $TeV$. The production rates depend on
NHL mass, $M_{N}$, and mixings. For the current limits on mixings, each of
the electron colliders is sensitive to values of $M_{N}$ approaching the
kinematic limit. The 500 $GeV$ muon collider is less sensitive because the
relevant mixing, $\mu\mu_{mix}$, is tightly constrained. For values of
$M_{N}$ below the kinematic limit, the various colliders are sensitive to
mixings in the range $1 \cdot 10^{-4}$ to $1 \cdot 10^{-6}$. Thus, the
process of single NHL production in lepton colliders is very promising for
the possible discovery of the NHL's. 

%

\end{document}